\documentclass[conference]{IEEEtran}
\IEEEoverridecommandlockouts

\usepackage{cite}
\usepackage{amsmath,amssymb,amsfonts}
\usepackage{algorithmic}
\usepackage{graphicx}
\usepackage{textcomp}
\usepackage{xcolor}
\usepackage{url,array}
\usepackage{multirow}
\usepackage{booktabs}
\def\BibTeX{{\rm B\kern-.05em{\sc i\kern-.025em b}\kern-.08em
    T\kern-.1667em\lower.7ex\hbox{E}\kern-.125emX}}
\begin{document}

\title{DiffAttack: Diffusion-based Timbre-reserved
Adversarial Attack in Speaker Identification
}

\author{
    \IEEEauthorblockN{
  \textit{Qing Wang}\IEEEauthorrefmark{2}, 
  \textit{Jixun Yao}\IEEEauthorrefmark{2}, 
        \textit{Zhaokai Sun}\IEEEauthorrefmark{2},
        \textit{Pengcheng Guo}\IEEEauthorrefmark{2},
        \textit{Lei Xie}\IEEEauthorrefmark{2}\IEEEauthorrefmark{1},
        \textit{John H.L. Hansen}\IEEEauthorrefmark{4},
  }
 \IEEEauthorblockA{\IEEEauthorrefmark{2}Audio, Speech and Language Processing Group (ASLP@NPU), School of Computer Science,\\ Northwestern Polytechnical University, Xian, China\\
 \IEEEauthorblockA{\IEEEauthorrefmark{4}Center for Robust Speech Systems (CRSS), The University of Texas at Dallas, USA
 }
}\thanks{\IEEEauthorrefmark{1} Corresponding author.}}

\maketitle

\begin{abstract}

Being a form of biometric identification, the security of the speaker identification (SID) system is of utmost importance. To better understand the robustness of SID systems, we aim to perform more realistic attacks in SID, which are challenging for humans and machines to detect. In this study, we propose DiffAttack, a novel timbre-reserved adversarial attack approach, that exploits the capability of a diffusion-based voice conversion (DiffVC) model to generate adversarial fake audio with distinct target speaker attribution. By introducing adversarial constraints into the diffusion-based voice conversion model's generative process, we aim to craft fake samples that effectively mislead target models while preserving the speaker-wised characteristics. Specifically, inspired by the utilization of randomly sampled Gaussian noise in conventional adversarial attack and diffusion processes, we incorporate adversarial constraints into the reverse diffusion process. As a result, these adversarial constraints subtly guide the reverse diffusion process toward aligning with the target speaker distribution. Our experiments on the LibriTTS dataset indicate that our proposed DiffAttack significantly improves the attack success rate compared to vanilla DiffVC or other methods. Furthermore, objective and subjective evaluations demonstrate that introducing adversarial constraints does not compromise the speech quality generated by the DiffVC model. 

\end{abstract}

\begin{IEEEkeywords}
diffusion model, adversarial attack, speaker identification, voice conversion.
\end{IEEEkeywords}

\section{Introduction}

Speaker identification~\cite{hansen2015speaker,reynolds1995robust} is an important area in speech processing and biometrics, aiming to recognize and verify individuals' identity based on their unique voice characteristics. Although this technology has been widely used in various fields, it is inevitably exposed to various attacks that affect its accuracy and reliability. Spoofing attacks~\cite{wu2012detecting, wu2015spoofing, wu2017asvspoof} typically replicate the timbre of the target speakers, commonly including impersonation, replay, voice conversion, and speech synthesis.
Whereas adversarial attacks~\cite{goodfellow2014explaining, carlini2017towards, szegedy2013intriguing, kurakin2016adversarial} confuse the SID system by introducing well-crafted perturbations into arbitrary speech. In recent years, many studies~\cite{das2020attacker,kreuk2018fooling, wang2019adversarial, abdullah2019practical,wang2020inaudible, li2020universal, xie2020real, li2020practical, li2020adversarial, chen2021real, jati2021adversarial,wang2023black} have proposed effective adversarial attacks on SID systems~\cite{snyder2018x, wan2018generalized, desplanques2020ecapa}.
However, while spoofing attacks imitate the timbre of the target speaker, they do not necessarily exploit the vulnerabilities of the SID model, and thus, they may not yield the desired prediction for the attacker. On the other hand, adversarial attacks can manipulate the SID system to produce a pre-determined decision but may not consistently align with specific text or speaker timbre requirements in certain attack scenarios. Consequently, qualified fake audio for attacking the SID model should be able to deceive both machines and humans simultaneously.

In our previous study~\cite{wang2023timbre}, we proposed a timbre-reserved adversarial attack in SID, in which we established a solution to first address the adversarial attack while maintaining the timbre and customizing the text. We added adversarial constraints during different training stages of the voice conversion (VC) models~\cite{fs2, kashkin2022hifi}. In~\cite{popov2021diffusion}, Popov \textit{et al.} proposed a voice conversion approach, denoted as DiffVC, based on diffusion probabilistic modeling (DPM)~\cite{ho2020denoising}, which is a scalable, high-quality solution with superior quality. 
DPMs have achieved significant attention due to their ability to capture complex data distributions and generate high-quality samples. 
In the context of DiffVC, the forward diffusion process gradually adds Gaussian noise to the data as an encoder. Conversely, the reverse diffusion process, which tries to remove this noise, acts as a decoder. In parallel to the Gaussian noise mentioned earlier, conventional adversarial attacks commonly utilize randomly sampled Gaussian noise in optimization. Drawing inspiration from this Gaussian noise in diffusion model and adversarial attack, for conducting more effective attacks in the SID system, we are considering whether we can use this adversarial constraint to implicitly guide the reverse diffusion process towards alignment with target speaker distribution.

To this end, in this study, we propose a novel approach, DiffAttack, to conduct timbre-reserved adversarial attacks in SID, incorporating adversarial constraints into DiffVC model training. 
In particular, an encoder and a decoder are served as forward and reverse diffusion processes, respectively. 
The adversarial constraints are integrated into the reverse diffusion process to guide the distribution closer to the target speaker. The adversarial constraints are more directly integrated into DiffVC model training instead of adding to the Mel-spectrograms in~\cite{wang2023timbre}.
Moreover, this strategy is applied to enhance the authenticity of the generated fake audio by better capturing the distinguishing attributes of the target speaker. 
We evaluate our proposed method on the LibriTTS dataset~\cite{zen2019libritts}. The fake audios generated by our proposed method improve the attack success rate significantly compared with the vanilla DiffVC model. The objective and subjective evaluations also illustrate that the quality of fake audio generated by our proposed method is better than other compared models.


\section{Methodology}\label{sec:method}


\subsection{System Overview}
\begin{figure}[h]
  \centering
  \includegraphics[width=8.8cm]{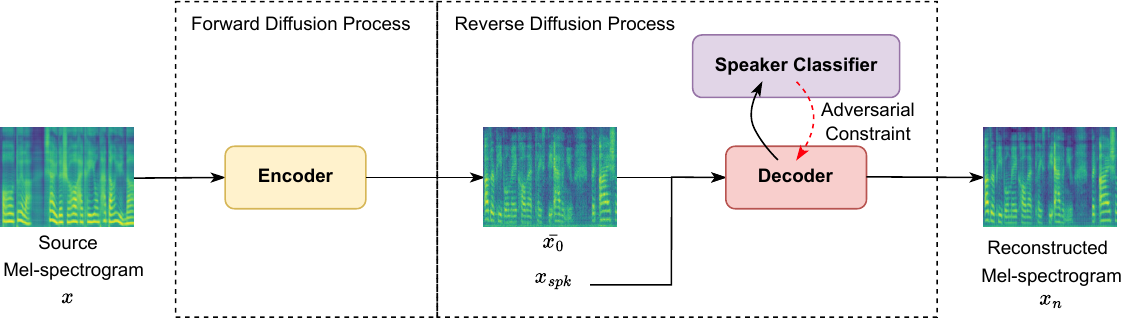}
  \caption{An overview of diffusion-based timbre-reserved adversarial attack in speaker identification.}
  \label{fig:overall}
\end{figure}

Our proposed approach's architecture is illustrated in Figure~\ref{fig:overall}, and it can be expressed as a conditional variational autoencoder. Specifically, our approach comprises two fundamental components: an encoder and a decoder, which serve as the forward and reserve diffusion processes, respectively. 
The encoder engages in a forward diffusion process, progressively introducing Gaussian noise into the data, while the decoder performs a reverse diffusion operation to eliminate this noise.

During the training phase, the encoder serves the purpose of generating a speaker-independent representation, and the decoder undertakes the task of reconstructing the target Mel-spectrogram, guided by the adversarial constraint. This adversarial constraint is employed to restrict the distribution of the reconstructed Mel-spectrogram, aligning it more closely with the distribution corresponding to the target speaker.

In the inference phase, the forward diffusion process uses the encoder to generate speaker-independent averaged Mel-spectrograms. Subsequently, the decoder is employed for the reverse diffusion process, with conditioning on both the timestep and the target speaker embedding. Following this, an independently trained HifiGAN vocoder~\cite{kong2020hifi} is applied to reconstruct waveform from Mel-spectrogram. As a result, timbre-reserved fake audio is generated and then utilized for carrying out adversarial attacks against the SID model. 

\subsection{Adversarial Diffusion Process}
\begin{figure}[h]
  \centering
  \includegraphics[width=8.8cm]{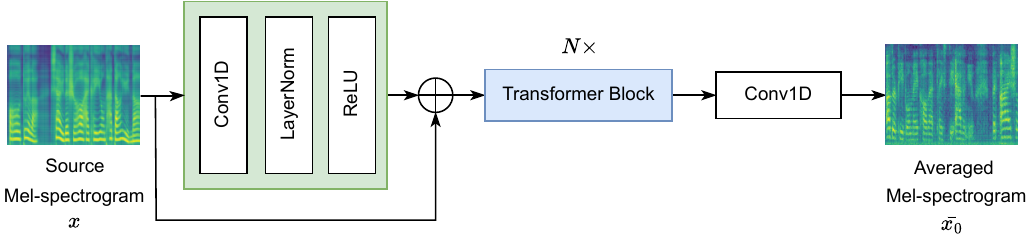}
  \caption{The architecture of the encoder, which is trained separately from the decoder.}
  \label{fig:encoder}
\end{figure}

In the forward diffusion process, the averaged Mel-spectrogram is employed as the target for the encoder. The primary training objective aims to minimize the mean square error (MSE) between the output of the encoder and the ground truth averaged Mel-spectrograms and is formulated as: 
\begin{equation}
    \mathcal{L}_{enc} = ||Enc(x)-\bar{x_0}||_2,
\end{equation}
where $Enc(\cdot)$ represents the encoder, $x$ and $\bar{x_0}$ represent the source and averaged Mel-spectrogram, respectively. This optimization objective ensures that the encoder learns to generate representations that closely match the target spectrogram during the forward diffusion.
As shown in Figure~\ref{fig:encoder}, the encoder follows the same architecture in Grad-TTS~\cite{popov2021grad}, with the distinction being that Mel-spectrogram features are used as inputs instead of characters or phonemes. 
It's worth noting that the encoder is trained separately from the decoder.

\begin{figure*}[h]
  \centering
  \includegraphics[width=15.8cm]{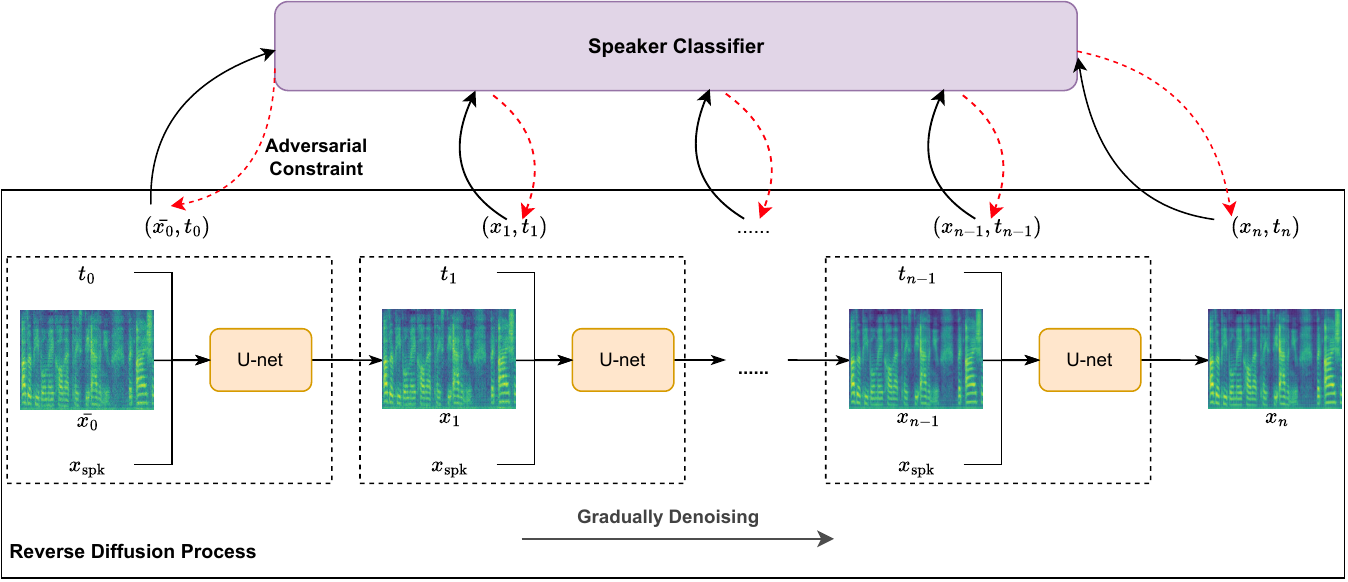}
  \caption{An overview of the reverse diffusion process. At each timestep $t$, the concatenation of speaker embedding $x_{spk}$ and averaged Mel-spectrogram $x_t$ is the input of the decoder. The speaker classifier then classifies $x_t$ determining if it is the target speaker to decide whether to add the adversarial constraint.}
  \label{fig:diff}
\end{figure*}

The reverse diffusion process begins with a disentangled speaker-independent representation and ends with the Mel-spectrogram that exhibits the desired timbre of the target speaker. To disentangle linguistic content information, average phoneme-level Mel-spectrogram features are employed, which are speaker-independent and are derived by aligning speech frames with phonemes through the Montreal Forced Aligner~\cite{mcauliffe2017montreal}. 
As shown in Figure~\ref{fig:diff}, during the reverse diffusion process, at each timestep $t$, we concatenate the speaker embedding $x_{\text{spk}}$ and the averaged Mel-spectrogram $x_t$ to form the input of the decoder. Meanwhile, it's worth noting that the encoder is responsible for parameterizing the terminal distribution of the forward diffusion, which serves as the prior distribution. Conversely, the decoder takes on the role of parameterizing the reverse diffusion process.

The decoder is designed as a part of a diffusion probabilistic model since this class of generative models has shown excellent performance in speech generation tasks, such as speech synthesis. In our approach, we formulate the reverse diffusion process in the context of stochastic processes. The DPM comprises both forward and reverse diffusion processes, with the reverse diffusion being employed by a decoder based on the U-Net architecture~\cite{ronneberger2015u}. Following stochastic differential equations (SDEs)~\cite{kloeden1992stochastic}, the forward diffusion process is calculated as:
\begin{equation}\label{eq2}
    d x_t=\frac{1}{2} \beta_t\left(\bar{x_0}-x_t\right) d t+\sqrt{\beta_t} d \vec{W}_t,
\end{equation}
where $\vec{W}_t$ and $\overleftarrow{W_t}$ are two independent Wiener process. Moreover, $t$ and $\beta_t$ are the diffusion timestep and non-negative functions referred to as noise schedules, respectively. The noise in each timestep follows the linear schedule and is formulated by:
\begin{equation}
    \beta_t=\beta_0+t\left(\beta_1-\beta_0\right).
\end{equation}
The reverse diffusion process is calculated as:
\begin{equation}
    d \hat{x}_t=\left(\frac{1}{2}\left(\bar{x_0}-\hat{x}_t\right)-s_\theta\left(\hat{x}_t, \bar{x_0}, t\right)\right) \beta_t d t+\sqrt{\beta_t} d \overleftarrow{W_t},
\end{equation}
where $s_\theta$ is the score function with parameters $\theta$.

Additional adversarial constraints are integrated into the reverse diffusion process to enhance the effectiveness of attacks in SID. In conventional adversarial attacks, optimization is typically performed using randomly sampled Gaussian noise, which aligns with the diffusion noise. In this case, the objective is for the adversarial constraints to implicitly guide the reverse diffusion process toward aligning with the target speaker distribution. This approach is designed to make the generated fake audio more convincing in learning the target speaker's distinctive characteristics.
As shown in the upper part of Figure~\ref{fig:diff}, the adversarial constraint is conducted by adding a well-trained speaker classifier. The predicted Mel-spectrogram is classified by the speaker classifier, and it is determined if it is the target speaker to decide whether to add the adversarial constraint. The averaged Mel-spectrogram $x_t$ at each timestep $t$ is the input of the speaker classifier, if the prediction of the speaker classifier is the target speaker, the model is optimized only using the original loss. In contrast, if the speaker classifier does not predict the target speaker label $y'$, an adversarial constraint $\delta$ is added to $x_t$, and can be optimized by: 
\begin{equation}\label{commo_method}
    \begin{aligned}
        & \min L_{\text{CE}}(f(x_t+\delta), y'), \text { s.t. } \quad \lVert \delta \rVert < \epsilon,        
    \end{aligned}
\end{equation}
where $L_{\text{CE}}(\cdot)$ aims to make the adversarial example ${x_t}_{adv} = x_t+\delta$ lead the SID model predicting the specified target label $y'$.
Subsequently, the joint training with adversarial constraint can be optimized by the following loss function: 
\begin{equation}
    \mathcal{L}_{\text{adv}} = ||{x_t}_{\text{gt}}-{x_t}_{\text{adv}}||_2,
\end{equation}
where $\mathcal{L}_{\text{adv}}$ is MSE loss between the ground truth and adversarial Mel-spectrogram at timestep $t$.

Consequently, the reverse diffusion is constrained by the adversarial speaker identification to minimize the $L_2$ loss:

\begin{equation}
    \begin{aligned}
        \theta^* = \arg\min_\theta & \int_0^1 \lambda_t \mathbb{E}_{x_0, x_t} \left\| s_\theta (x_t, \bar{x}_0, t) \right.  \\
        & \left. - \nabla \log p_{t|0} (x_t | \bar{x}_0) \right\|_2^2 dt, 
    \end{aligned}
\end{equation}

where $p_{t \mid 0}\left(x_t \mid \bar{x_0}\right)$ is the probability density function of the target speaker conditional distribution and $\lambda_t=1-e^{-\int_0^t \beta_s d s}$. During the training phase, timestep $t$ is sampled uniformly from $[0,1]$, and noise samples $x_t$ are obtained by Equation~\ref{eq2}.

\section{Experimental Setup}
\label{sec:setup}


\subsection{Datasets}
In this study, LibriTTS~\cite{zen2019libritts} corpus is used to train the diffusion-based VC model, which is a multi-speaker English corpus of approximately 585 hours of read English speech and is derived from the original materials of the LibriSpeech corpus. The Voxceleb 1\&2~\cite{nagrani2017voxceleb, chung2018voxceleb2} corpora are used to train the SID model, and we use the LibriTTS corpus to finetune the SID model. The test set of Voxceleb 1-O is used to evaluate the performance of the SID model.
For the test set, we randomly select 500 utterances from the LibriTTS corpus as the source speech. 10 target speaker labels (5 female speakers and 5 male speakers) are selected from the LibriTTS corpus, which are different from the speaker labels of source speech. Then we use the source speech converted into these 10 target speakers' speech and we test the performance of the fake audio.
\subsection{Setup}
The detailed experimental setup of all the models used in this study is described as follows: ECAPA-TDNN~\cite{desplanques2020ecapa} is used as the SID model and also used as the speaker classifier in the diffusion VC model. The EER of this model on the Voxceleb 1-O test set is 1.49\%. We finetune the ECAPA-TDNN(c=1024) model on the 16k LibriTTS dataset using the weight transfer method in~\cite{zhang2023distance}. 
The setup of the vanilla DiffVC model in this study is according to the configuration in ~\cite{popov2021diffusion}. HifiGAN vocoder~\cite{kong2020hifi} is followed to reconstruct the waveform from Mel-spectrogram.  
As a compared model, we add a speaker classifier directly to the vanilla DiffVC model to constrain the Mel-spectrogram in the reverse diffusion process, denoted as `DiffVC+SPK Constrain'.
As shown in Figure~\ref{fig:diff}, the speaker classifier is added to identify whether the Mel-spectrogram of timestep t is the target speaker or not and is denoted as `DiffVC+ADV Constrain'. 


\section{Experimental Results}
\label{sec:result}

\subsection{Attack Success Rate}
As shown in Tabel~\ref{tab:Acc}, `Acc' is the attack success rate, used to measure the performance of targeted attacks in SID. It represents the accuracy inferred from the SID and is calculated by dividing the number of fake audio successfully attacking the model by the total number of fake audio samples. A higher 'Acc' signifies a superior attack performance.
The baseline `DiffVC' generates fake audio by the vanilla DiffVC model. `DiffVC+ADV Perturb' is the upper limit that directly adds adversarial perturbation to the fake audios generated by the DiffVC model, using the approach proposed in~\cite{wang2020inaudible}. 
Another compared method is proposed in~\cite{wang2023timbre} denoted as `VC+ADV Constraint', generating fake audio by adding adversarial constraint during the end-to-end VC model training.
The remaining two methods (`DiffVC+SPK Constraint' and `DiffVC+ADV Constraint') are based on the DiffVC model trained with speaker classifier and adversarial constraints.

As shown in Table~\ref{tab:Acc}, the attack success rate of fake audio generated by vanilla DiffVC model-based is 28.40\%, while the DiffVC audio with direct adversarial perturbation~\cite{wang2020inaudible} achieves 73.40\%. 
Compared to the vanilla DiffVC model, the attack success rate of fake audio generated through our proposed method experiences a notable enhancement, with an improvement of 37.36\%.
Furthermore, the `DiffVC+ADV Constraint' method outperforms our prior work~\cite{wang2023timbre} by 8.46\%. This improvement can be attributed to introducing the adversarial constraint during the diffusion process, rendering it more delicate than the approach presented in~\cite{wang2023timbre}.
The proposed method improves by 8.16\% compared to `DiffVC+SPK Constraint', which indicates that adversarial constraints are more effective than the direct use of speaker classifier constraints.

\vspace{-8pt}
\begin{table}[h]\centering
\caption{Attack success rates (\%) of different generation methods.}
\label{tab:Acc}
\renewcommand{\tabcolsep}{0.45cm}
\renewcommand\arraystretch{1.5}
\begin{tabular}{ccc}
\toprule
                  & Method   & Acc (\%) $\uparrow$ \\ \hline
Baseline          & DiffVC        & 28.40   \\
Upper limit       & DiffVC+ADV Perturb~\cite{wang2020inaudible}    & 73.40   \\ \hline
                  & VC+ADV Constraint~\cite{wang2023timbre} &   57.30  \\
                  & DiffVC+SPK Constraint &   57.60 \\  
\textbf{DiffAttack} & \textbf{DiffVC+ADV Constraint} &   \textbf{65.76} \\                  
                  \bottomrule
\end{tabular}
\end{table}
\vspace{-8pt}

\subsection{Objective and Subjective Evaluation}
\subsubsection{Objective evaluation}
\begin{figure}[h]
  \centering
  \includegraphics[width=8.5cm]{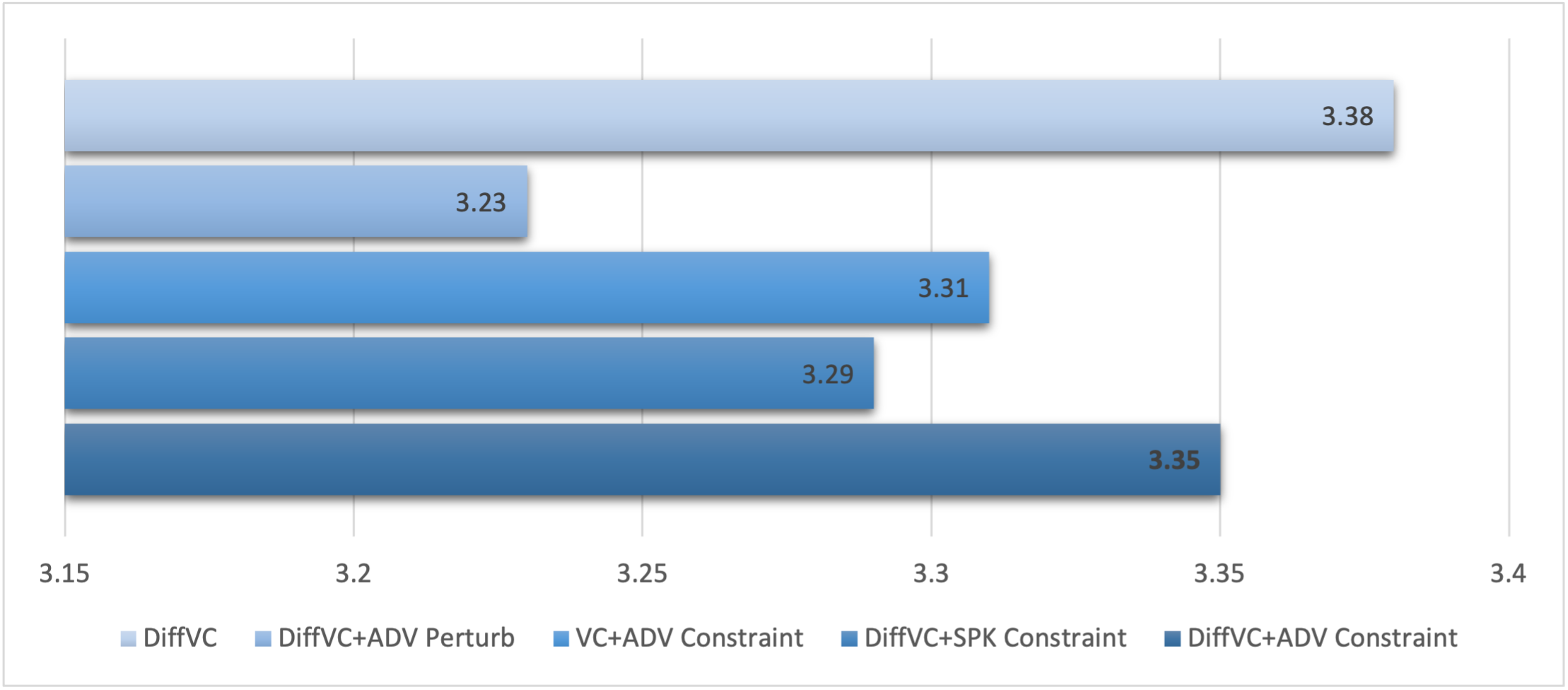}
  \caption{The o-MOS of different kinds of generation methods.}
  \label{fig:obj}
\end{figure}

The MOSNet~\cite{lo2019mosnet} prediction o-MOS is employed as the objective measurement and Figure~\ref{fig:obj} shows the objective performance of the o-MOS of various generation methods. 
We can observe that the o-MOS results for fake audio generated through the constraint-based strategies consistently surpass those obtained by directly incorporating adversarial perturbations into the audio since we avoid directly introducing extra perturbations to the fake audio.

\subsubsection{Subjective evaluation}
We conduct a CMOS evaluation, as presented in Table~\ref{tab:obj}. We can observe that the results of these three constraint-based methods outperform the `DiffVC+ADV Perturb' method. 
This demonstrates that fake audio generated using these three constraint-based methods closely resembles audio generated by the vanilla DiffVC model, and the constraint-based strategies effectively mitigate the influence of this extra perturbation by incorporating adversarial constraints into the generation model. 

Furthermore, MOS tests are also conducted on various generation methods to assess the quality and speaker similarity of the generated fake audio. 
Regarding audio quality, our proposed strategy outperforms all other compared methods. In comparison to the `VC+ADV Constraint'~\cite{wang2023timbre} method, our proposed method is improved, since the adversarial constraint added in the diffusion-based VC model is more refined than the traditional VC models.
Regarding speaker similarity, the MOS results for our proposed strategies also slightly surpass those of the other two constraint-based methods. These subjective results indicate that our diffusion-based adversarial timbre-reserved fake audio effectively maintains the VC model's capabilities and can better characterize the attribute of the target speaker.

\vspace{-8pt}
\begin{table}[h]\centering
\caption{CMOS and MOS results of fake audios by different strategies.}
\label{tab:obj}
\renewcommand{\tabcolsep}{0.25cm}
\renewcommand\arraystretch{1.3}
\begin{tabular}{cccc}
\bottomrule
\multirow{2}{*}{} & \multirow{2}{*}{CMOS $\uparrow$} & \multicolumn{2}{c}{MOS $\uparrow$} \\ \cline{3-4} 
                  &                       & Quality   & Similarity  \\ \hline
DiffVC                & -                     & 3.92$\pm0.07$      & 3.88$\pm0.06$        \\
DiffVC+ADV Perturb ~\cite{wang2020inaudible}          & -0.35                 & 3.40$\pm0.06$      & 3.71$\pm0.05$        \\
\hline
VC+ADV Constraint~\cite{wang2023timbre}          & -0.18                 & 3.65$\pm0.04$      & 3.76$\pm0.06$        \\
DiffVC+SPK Constraint           & -0.17                 & 3.72$\pm0.04$      & 3.74$\pm0.04$        \\
DiffVC+ADV Constraint           & -0.13                 & \textbf{3.88$\pm0.05$}      & \textbf{3.85$\pm0.04$}        \\ \bottomrule
\end{tabular}
\end{table}
\vspace{-6.3pt}

\section{Conclusion}
\label{sec:conclu}

In this study, we propose DiffAttack to generate timbre-reserved fake audio using a DiffVC model for adversarial attacks on SID systems. By adding adversarial constraints to the reverse diffusion process, we create fake audio that retains both timbre and speaker-specific characteristics, effectively attacking the SID model. Our experiments on the LibriTTS corpus show that our strategy significantly improves the attack success rate compared to other generation methods. Objective and subjective evaluations reveal that adversarial constraints do not compromise the fake audio quality and outperform other kinds of constraints.

\end{document}